\documentstyle[12pt]{article}
\def \dst {\displaystyle}

\font\goth=eufm10 scaled 1400
\font\gotic=eufm10 scaled 1100

\newcommand{\lt} {\ {\bf <}}
\newcommand{\rt} {{\bf >}\ }
\newcommand{\lb} {\left(}
\newcommand{\rb} {\right)}
\newcommand{\ld} {\left.}
\newcommand{\rd} {\right.}
\newcommand{\cross} {\mbox{\hspace{-5 mm} / \hspace{1 mm}}}
\newcommand{\al} {\alpha}
\newcommand{\add} {\addtocounter{prop}{1}}

\newcommand{\be} {\beta}
\newcommand{\cf} {(n-1)}

\newcommand{\ga} {\gamma}
\newcommand{\bga}{\begin{array}{l}}
\newcommand{\ena}{\end{array}}
\newcommand{\bge}{\begin{equation}}
\newcommand{\ene}{\end{equation}}
\newcommand{\prp}{\add \noindent {\it Proposition \theprop .} \\}
\author{ M. Zyskin  \thanks{IHES, Le Bois-Marie 35, route de Chartres
F-91440, Bures-sur-Yvette, France; zyskin@ihes.fr} }
\title{Non-highest weight representations of the current algebra
$\widehat{so}(1,n)$,
and Laplace Operators.}

\date{March, 1999}

\begin{document}
\maketitle
\vspace{-200pt}
hep-th
\vspace{200pt}

\begin{abstract}
We   constructed canonical non-highest weight unitary irreducible
representation of $\widehat{so}(1,n)$ current algebra as well as canonical
non-highest weight non-unitary representations,
We constructed certain Laplacian operators as elements of the universal
enveloping algebra, acting in representation space. We speculated about a
possible relation of those Laplacians with the loop operator for the
Yang-Mills.
\end{abstract}

\newpage

\section{Introduction}

Recently, there was a lot of interest in  string theories on AdS and  other
manifolds with boundaries, and to their relationship to Yang Mills quantum
field theories
on the boundary.
String theory on AdS has a symmetry group, corresponding to isometries of
target space, whose bosonic part is $SO(1,n)$.
There should be   conserved Noether currents corresponding to such
symmetries.
Also, such string theories should be described by conformal field theories.
If the conformal field theory is unitary, the currents are holomorphic, and
give rise
 to a Kac-Moody algebra. From analyticity on world sheet, the space of states
in such
conformal
theory, associated to points on worldsheet, should be in highest  weight
modules  of the Kac Moody algebra, and then there is a
Virasoro algebra
coming  via the Sugawara construction \cite{kz}. Namely, the currents are
(closed)
one forms on
the world sheet ${{\mbox{{\gotic j}}^a}(.)}_\mu dx^{\mu}$, the integrals of
which over paths
gives conserved charges, with values in Lie algebra; and "equal time"
commutation
relations of currents should be such as to reproduce the Lie algebra
commutators of charges; if
we integrate the current over a path $\Gamma$ on world sheet with  weights
$\psi(\ga)$, $\ga \in \Gamma$,
 and use commutation relations for currents, we will get
$$ [ \dst\int_\Gamma \  \psi \ \mbox{{\gotic j}}^a d \ga ,
\dst\int_\Gamma \  \xi \ \mbox{{\gotic j}}^b d \ga ]= f^{abc} \dst\int\
\psi \ \xi \  \mbox{{\gotic j}}^c d \ga + \tilde{k} \dst\int \psi^\prime \xi
d\ga; $$
and in particular, taking a closed path, parametrised by $\theta$,  and $\psi,
\xi$
of the form $e^{i n \theta}$, we get that Fourier coefficients of currents
satisfy
commutation relations of a Kac-moody algebra.

In our case, the algebra is $so(1,n)$, which is  a  real form of a complex
algebra;
still it can be
promoted to a Kac-Moody-like loop
algebra in the usual way; it is generated by
$
e_i\otimes_{\bf R} P(s)$, with $e_i \in so(1,n)$, $P(s)$ Laurent polynomials,
and the
bracket
$
[e_i\otimes P(s),e_j\otimes G(s)] = [e_i, e_j]\otimes P(s) G(s) + \tilde{k}
\lb
e_i, e_j \rb \ Res  ( \frac{dP}{ds}  G ) ,
$
where $\lb . , . \rb$ is an invariant symmetric  form on $so(1,n)$; this
form is not positive-definite.
Highest weight module for such  loop algebra is not unitary.

If we think about boundary states, those associated with the boundary of world
sheet disk, say states associated to a circle, we cannot deduce from
analyticity that the state
must be highest weight, as functions $\{ e^{i n \theta} \}$ on a circle, unlike
functions $\{ z^n \}$
at the origin,
are equally good for positive or negative
n,  and in fact
there are two possible candidates
for representations: one non-unitary  highest weight,
and another unitary, not highest weight. Non highest weight representations are
little studied
in the framework of the conformal field theory; in particular, there is no
meaningful  way
to have a Sugawara construction of the Virasoro algebra; instead, it is quite
natural to get
certain infinite dimensional laplacians in the universal enveloping algebra.

\section{Unitary irreducible representation of  $\widehat{so}(1,n)$ loop
algebra and unitary action of  operators in universal enveloping algebra}
\newcounter{prop}
\setcounter{prop}{0}
\renewcommand{\theprop}{\thesection.\arabic{prop}}
\subsection{}
We introduce a space {\goth s} of real valued functions on a circle, which is a
vector space over real numbers
with basis ${\{e[n]\equiv s^n\ \} }_{n\in \bf{Z}}$, and with some
positive definite inner product
\bge
\bga
(e[n], e[m]) =F[n,m]\\
F[n,m]=F[m,n]; F[n,m]\in {\bf R^{+}},
\ena \label{inners}
\ene
extended by linearity to the whole space.
Usual multiplication of functions give rise to a product, $e[n]\cdot e[m]
=e[m+n]$. If we  want multiplications by $s \equiv e[1]$ and $s^{-1} \equiv
e[-1]$ to be self-adjoint operators
$$
\bga
(e[n], s\cdot e[m])= ( s \cdot e[n], e[m])\\
(e[n], s^{-1} \cdot e[m])= ( s^{-1} \cdot e[n], e[m]),
\ena
$$
then from (~\ref{inners}) we have a condition
$$
F[n,m+1] = F[n+1,m].
$$
Therefore in this case we should have
\bge
F[n,m] = \tilde{F} [n+m, n-m \ mod \  2]
\label{unitary}
\ene
where $n\  mod \  2 $ is 0 for even n and 1 for odd n. Such scalar products
exist:\\
{\it Example:} consider the inner product
$$
(e[n], e[m]) = \dst\int_{- \infty}^{\infty} t^n \ t^m \ Exp[-(t^2 + 1/t^2 )] \
dt
$$
It is obviously positive definite and $(e[n], e[m])$ depends only from $n+m$

We can and will consider other scalar products where mutiplication by $s$ is
not self-adjoint.\\
{\it Example}: intoduce a sclar product
\bge
(e[n], e[m]) = \delta_{n,m} \ \al (n)
\label{nonun}
\ene
where $\al (n) >0$ for all $n\in {\bf Z}$. Such scalar product is obviously
positive-definite.
{}From the definition of the adjoint operator $(e[n], s\ e[m])=(s^* \ e[n],
e[m])$is easily follows that in this case
$$
s^* \ e[n] =  \dst\frac{\al (n)}{\al (n-1) } \ e[n-1]
$$

\subsection{}
The loop algebra $\widehat{so}(1,n)$ has an interesting unitary,
non-highest weight irreducible representation, which we believe is
important and which we describe below.
Consider the Hilbert space {\gotic h } of real valued functions $\phi(X)$ on a
sphere $ S^n$: ${X_1}^2 +
{X_2}^2 +\ldots + {X_n}^2 =1$ with
integral zero, $\dst\int \phi(X) d\omega_X =0$, and with the scalar product
\bge
\langle \phi , \psi \rangle =
\dst\int\dst\int \phi (X) \psi(X^\prime) \ln \left( 1- (X, X^\prime ) \right)
d\omega_X d\omega_{X^\prime}
\ene
It is possible to show that for functions on a sphere with zero integral such
scalar product is
positive- definite.

Let us  introduce a Hilbert space {\goth H} of maps $\phi (X,s)$ from $S^1$
into {\gotic h}. For Laureant polynomials,
$$
\phi (X,s) =\dst\sum_{n=-N}^{M} \phi_n  (X) \ e[-n] (s)$$
where $ \phi_n  (X)\in  {\gotic h}$, and $\{ e[n]\}_{n\in {\bf Z}} $ is the
basis  in {\goth s}.
we define the scalar product in {\gotic H}
 as follows:
$$
\lt \phi () , \psi ()\rt = \dst\sum_{m,n}  (e[-n], e[-m])  \ \dst\int\dst\int
\phi_n (X)
\psi_m (X^\prime) \ln \left( 1- (X, X^\prime ) \right)
d\omega_X d\omega_{X^\prime},
$$
and then take the completion.

We introduce also  bosonic Fock space {\goth F} of  symmetric tensor products
{\goth F} $= \bigoplus_{m} S \ \mbox{{\goth H}}^{\otimes m}   $,
which is also a  Hilbert space with the scalar   product
\bge
\bga
\langle  {\phi ()}^{\otimes m}\ , \
{\psi ()}^{\otimes m}  \rangle
=
 m!\  {\langle  {\phi ()} \ , \
{\psi ()}  \rangle }^m
\ena
\ene
or more generally,
\bge
\bga
\lt \eta_1 \otimes \eta_2 \ldots  \otimes \eta_n ,
\xi_1 \otimes \xi_2 \ldots \otimes \eta_m  \rt = \\
= \left\{
         \bga  0,\ \ \  m\neq n \\
             \dst\sum_{\{ \sigma \}} \ \dst\prod_{i=1}^{n}  \lt \eta_{i} ,
\xi_{\sigma_i} \rt ,\ m=n
         \ena
  \right.
\ena \label{scalarF}
\ene

We also introduce  $Exp (\phi())$ in a completion of {\goth F},
\bge
Exp (\phi(X, s)= 1+ \sum_{m=1}^{\infty} \ \dst\frac{1}{m!}\  {\phi
(X, s)}^{\otimes m}
\ene
It's easy to see that
\bge
\bga
\langle Exp( \phi()) , Exp (\phi ()) \rangle = exp \left( \langle \phi(), \phi
() \rangle \right)
\ena
\ene

\subsection{Unitary irreducible representation of loop algebra
$\widehat{so}(1,n)$ via first order
differential operators acting in  {\goth F} }
We consider maps   $S^1 \rightarrow $ {\gotic g} from a circle into the Lie
algebra $so(1,n)$, with the
point-wise Lie bracket,
\bge
\left\{I_1, I_2\right\} (s)= \left\{ I_1 (s), I_2(s)\right\}
\ene
This Lie algebra $\widehat{so}(1,n)$ is given by generators and relations
\bge
\bga
\left\{ I_{ij} ( f(s)), I_{kl} (g(s))\right\}  = \delta_{jk} I_{il} (fg (s))
-\delta_{jl} I_{ik}(fg (s)) +
\delta_{il} I_{jk} (fg (s)) -\delta_{ik} I_{jl} (fg (s))
\\[5mm]
\left\{ I_{ij} ( f(s)), I_{k,0} (g(s))\right\}  = \delta_{jk} I_{i,0} (fg
(s))-\delta_{ik} I_{j,0} (fg (s))
\\[5mm]
\left\{ I_{i,0} ( f(s)), I_{j,0} (g(s))\right\}  = I_{i,j} (fg (s))
\ena
\ene
This algebra has the following representation in {\goth F}:
for any  $v= \eta_1 \otimes \eta_2 \otimes \ldots \otimes \eta_n \in
\mbox{{\goth F}}$,
with  $\{ \eta_i \}  \in \mbox{\goth H}$,
\bge
\bga
T[I  (f(s))] . v  = \Bigl( D[I[f(s)]] + \be [I[f(s)]] -\  \lt \be [I[{f(s)}^*
\ 1]], \Bigr). v \equiv \\
\equiv \dst\sum_{i}    \eta_1 \otimes \eta_2 \otimes \ldots \otimes \bigl(
D[I[f(s)]]\eta_i  \bigr)
\ldots \otimes \eta_n +
 \be [I[f(s)]] \otimes \eta_1 \otimes \eta_2 \otimes \ldots \otimes \eta_n - \\
- \dst\sum_{i} \lt \be [I[f(s)^* 1]], \eta_i \rt\  \eta_1 \otimes \eta_2
\otimes \ldots \eta_i \mbox{\hspace{-4.5 mm} /
 \hspace{1 mm}} \ldots \otimes \eta_n
\ena \label{rule}
\ene

Here $D$ are derivations, which satisfy the Leibnitz rule in {\goth F},
$$
D[I( f(s))] \ \eta_1 \otimes \eta_2 \otimes \ldots \otimes \eta_n
=\dst\sum_{k}
\eta_1 \otimes \eta_2 \otimes \ldots \otimes \Bigl( D[I( f(s))] \eta_k \Bigr)
\otimes \ldots \otimes \eta_n
$$
and are given by the following differential operators in {\goth H}:

\bge
\bga
D[I_{ij}(f(s))] \ \phi (X, s) = f(s) \left( X_i \partial_j -X_j
\partial_i)\right) \phi(X,s),
\\[7mm]
D[I_{i,0} (f(s))] \ \phi (X, s) =f(s) \left( \partial_i - X_i (X \partial) -
\cf X_i \right) \phi(X,s),
\\[4mm]
D[I_{i,0} (f(s))] \ 1 = 0;
\ena
\label{D}
\ene
here $\partial_i \equiv \dst\frac{\partial}{\partial_i}$ in ${\bf R}^{n}$, so
that
$[\partial_i, X_j ]= \delta_{i,j} $;
\vspace{5mm}

\noindent and $\be$ acts via tensor multiplication in {\goth F}, with
\bge
\bga
\be [I_{ij}(f(s))] =0
\\[4mm]
\be [I_{i,0}(f(s))]   = -  \cf f(s) \xi_i (X),
\ena
\ene
where $\xi_i(X)$ is a function on a sphere equal to the i-th coordinate,
$\xi_i(X)= X_i $.
\\[4mm]
Such $\be$ is a one-cocycle:
\bge
D[ I_1] \be [ I_2] - D[ I_2] \be [ I_2] = \be[ \ \left\{ I_1, I_2\right\} \  ]
\ene

\add \noindent {\it Proposition \theprop .} \\
1) Differential operators $D[I(f(s))]$ are correctly defined  in {\goth H} \\
2)The operators $D [I(f(s))]^*$, which are adjoint in {\goth H} to operators $D
[I(f(s))]$,
 are equal to $(- D [I( {f(s)}^*  )]$,
where $f^* (s)$ is the adjoint operator to the multiplication by $f(s)$ in
{\goth s}
$$
D [I(f(s))]^* = (- D [I( {f(s)}^* )]);
$$
in particular,
if multiplication by $s$ is self  adjoint in {\goth s}, see
(~\ref{unitary}), then $D [I(f(s))]$ are   scew self-adjoint in {\goth H}.

Operators $D [I( {f(s)}^* )]$  act naturally in {\goth H}:
if $\eta = \sum \eta_n g_n$, with $\{ \eta_n \} \subset \mbox{\goth h}$, $\{
g_n \} \subset \mbox{\goth s}$, then
$$
D [I( {f(s)}^* )] \ \eta = \dst\sum D[I] \eta_n \ \ {f(s)}^* g_n,
$$
where $D[I]$ are the differential operators (~\ref{D}).
\vspace{5 mm}

\add \noindent {\it Proposition \theprop}\\

1) Operators $T[I  (f(s))] $ in {\gotic F} have the property $T[I  (f(s))]^* =
- T[I  ({f(s)}^* \ 1)] $; in particular,
if multiplication by $s$ is self  adjoint in {\goth s}, see
(~\ref{unitary}), then $T[I  (f(s))] $ are   scew self-adjoint in {\goth F}. \\
2) Operators $T[I  (f(s))] $ satisfy the commutation relation of the algebra,
that is for any $v\in \mbox{{\goth F}} $
$$
\Bigl( I_1 [f_1] I_2 [f_2] - I_2 [f_2] I_1 [f_1] \Bigr). v = \Big[ I_1, I_2
\Bigr] [ f_1\cdot f_2]. v
$$
{\it Proof}:
to prove 1), use (~\ref{rule}) and (~\ref{scalarF})
to show that  that for any $v= \eta_1 \otimes \eta_2 \otimes \ldots \otimes
\eta_n$ and
$w = \xi_1 \otimes \xi_2 \otimes \ldots \otimes \xi_n$
in $ S^n \mbox{{\goth H}}$

$$
\bga
\lt w , D[I[f(s)]]. v \rt =
\lt w, \dst\sum_{i}    \eta_1 \otimes \eta_2 \otimes \ldots \otimes \bigl(
D[I[f(s)]]\eta_i  \bigr)
\ldots \otimes \eta_n \rt =
\\
= \dst\sum_{\{ \sigma \} }  \dst\sum_i \lt \xi_{\sigma (i)},  D[I[f(s)]]\eta_i
\rt
\dst\prod_{j\neq i} \lt \xi_{\sigma (j)}, \eta_j \rt =
\\
= \dst\sum_{\{ \sigma \} }  \dst\sum_i \lt \bigl( - D[I[f^* (s) ]] \bigr)
\xi_{\sigma (i)}, \eta_i   \rt
\dst\prod_{j\neq i} \lt \xi_{\sigma (j)}, \eta_j \rt =
\\
= \lt \bigl( - D[I[f^* (s) ]] \Bigr) .   w , v \rt
\ena
$$
Since in {\goth F}  $\lt S^n \mbox{\goth H} , S^m \mbox{\goth H} \rt $ is zero
for $m\neq n$, the above equality
is true for any $v,w \in \mbox{\goth F}$
Similarly, since for any $v= \eta_1 \otimes \eta_2 \otimes \ldots \otimes
\eta_{n-1}$ in $ S^{n-1} \mbox{{\goth H}}$
 and
$w = \xi_1 \otimes \xi_2 \otimes \ldots \otimes \xi_n$
in $ S^n \mbox{{\goth H}}$,
$$
\bga
\lt w , \be [I(f(s))]  \otimes v \rt = \dst\sum_{\{ \sigma \} }  \dst\sum_i \lt
\xi_{\sigma (i)},  \be  [I(f(s))] \rt
\dst\prod_{j\neq i} \lt \xi_{\sigma (j)}, \eta_j \rt =
\\
\lt \ \ \Bigl( < \be [I[f(s)]], \Bigr). \ w  ,\  v  \ \rt ,
\ena
$$
and   also for any $v \in S^{n} \mbox{{\goth H}}$
 and
$w  \in  S^{n-1} \mbox{{\goth H}}$,
$$
\lt \ w  ,\ \ \Bigl(  - < \be [I[f^* (s)\ 1]], \Bigr). \  v  \ \rt =
\lt \ - \be [I[f^* (s)\ 1]] \otimes w , \ v \ \rt
$$
Collecting all terms, we obtain 1)\\

2) can be verified by a straitforward computation

\subsection{Representation of the Universal Enveloping Algebra in {\goth F} }
Elements in the universal enveloping algebra, which are polynomials in
generators, are represented by operators acting
in {\goth F} given by corresponding polynomials of operators $T[I  (f(s))] $.
The commutation relations of the algebra
are satisfied due to Proposition.

\subsection{Operators $\bf L_m$}
\noindent {\it Definition: operators $\bf L_m$}

Define   $\bf L_m$ to be elements of  the universal enveloping algebra
given by
\bge
\bga
\mbox{{\bf L}}_m =
  \lb \dst\sum_{k = -\infty}^{+\infty} \dst\sum_{i<j}  T[ I_{ij}(e[-k] )]\
 T[I_{ij}( e[k+m] )] \rd
- \\
- \ld \dst\sum_{i} T[ I_{i,0}( e[-k] ) ]\  T[ I_{i,0}(e[k+m)] ]\rb - C(m) {\bf
1},\\
C(m) = \dst\sum_{k = -\infty}^{+\infty}
 \dst\sum_{i}\lt \be [I_{i,0}(e[-k]^* \ 1)] \ , \be[I_{i,0}( e[k+m] )]
 \rt
\ena
\label{l}
\ene
where operators $T [I (f(s)]$ are defined in (~\ref{rule}), and $\{ e[k]\}$ is
the basis  in {\goth s},
see (~\ref{inners}).
\\[6mm]

\add \noindent {\it Proposition \theprop .} The action of   ${\bf L}_m$ in
{\gotic F}
 as  a formal series is well defined.
\\[4 mm]
Indeed,
it is easy to check that for an $\eta \in \mbox{{\goth H}}$ and any $k,m \in Z$
$$
\Bigl( \dst\sum_{i<j}   D[I_{ij}(e[-k] )]\
 D[I_{ij}( e[k+m] ) ]
- \dst\sum_{i}  D[I_{i,0}( e[-k] )] \   D[I_{i,0}(e[k+m])] \Bigr) \ \eta = 0
$$
and
$$
\dst\sum_{i}  D[I_{i,0}( e[-k] )] \ .  \be [I_{i,0}(e[k+m])] = 0
$$
Therefore, dangerous terms disappear, and for  any $v = \eta_1 \otimes \eta_2
\otimes \ldots \otimes \eta_n \in \mbox{{\goth F}}$,
$$
\bga
{\bf L}_m .v =  \dst\sum_k \Bigl(
\\
\dst\sum_{i<j,a,b} \eta_1 \otimes \eta_2 \otimes \ldots \otimes D[I_{i,j}(
e[-k] )]\eta_a \otimes \ldots \otimes D[I_{i,j}(e[k+m])]\eta_b \otimes \ldots
\otimes \eta_n -
\\
-\dst\sum_{i,a,b} \eta_1 \otimes \eta_2 \otimes \ldots \otimes D[I_{i,0}( e[-k]
)]\eta_a \otimes \ldots \otimes D[I_{i,0}(e[k+m])]\eta_b \otimes \ldots \otimes
\eta_n -
\\
 - \dst\sum_{i,a} \eta_1 \otimes \eta_2 \otimes \ldots \otimes D[I_{i,0}( e[-k]
)]\eta_a \otimes  \ldots \otimes \eta_n  \otimes \be [I_{i,0}(e[k+m])] -
\\
- \dst\sum_{i,a} \eta_1 \otimes \eta_2 \otimes \ldots \otimes D[I_{i,0}( e[k+m]
)]\eta_a \otimes  \ldots \otimes \eta_n  \otimes \be [I_{i,0}(e[-k])]-
\\
- \dst\sum_{i} \eta_1 \otimes \eta_2 \otimes \ldots  \otimes \eta_n  \otimes
\be [I_{i,0}(e[-k])]\otimes \be [I_{i,0}(e[k+m])]+
\\
+\dst\sum_{i,a,b}\lt \be [I_{i,0}(e[-k]^* \ 1 )] \ ,\eta_a \rt \  \eta_1
\otimes \eta_2 \otimes \ldots \otimes D[I_{i,0}( e[k+m] )]\eta_b \otimes \ldots
\otimes \eta_a \cross \otimes \ldots \otimes \eta_n
\\
+\dst\sum_{i,a,b}\lt \be [I_{i,0}(e[k+m]^* \ 1 )] \ ,\eta_a \rt \  \eta_1
\otimes \eta_2 \otimes \ldots \otimes D[I_{i,0}( e[-k] )]\eta_b \otimes \ldots
\otimes \eta_a \cross \otimes \ldots \otimes \eta_n
\\
+\dst\sum_{i,a}\lt \be [I_{i,0}(e[-k]^* \ 1 )] \ ,\eta_a \rt \  \eta_1 \otimes
\eta_2 \otimes \ldots \otimes \ldots \otimes \eta_a \cross \otimes \ldots
\otimes \eta_n \otimes \be[I_{i,0}( e[k+m] )]
\\
+\dst\sum_{i,a}\lt \be [I_{i,0}(e[k+m]^* \ 1 )] \ ,\eta_a \rt \  \eta_1 \otimes
\eta_2 \otimes \ldots \otimes \eta_a \cross \otimes \ldots \otimes \eta_n
\otimes \be[I_{i,0}( e[-k] )]-
\\
-\dst\sum_{i,a,b} \lt \be [I_{i,0}(e[-k]^* \ 1 )] \ ,\eta_a \rt \
\lt \be [I_{i,0}(e[k+m]^* \ 1 )] \ ,\eta_b \rt \  \\
\hspace{94 mm} \eta_1 \otimes \eta_2 \otimes \ldots \otimes
\eta_a \cross \otimes \ldots \otimes \eta_b \cross
\otimes \ldots \otimes \eta_n
\ena
$$

\noindent {\it Remark}.  It is easy to check, that unlike in Sugawara
construction of the Virasoro
algebra from current algebra in highest weight modules, here a "normal
ordering"
of operators $T[I]$ in $L$ gives the same expression as the one without the
normal ordering.
\vspace{7mm}

\prp
1) $L_m$ commutes with all operators $T[I(f(s))]$, with $f(s) \in \mbox{\goth
s} $\\
2.a)For scalar product in {\goth s} such that multiplication by s is a unitary
operator, (~\ref{unitary}),
operators $L_m$ are formally ( as a formal series) self-adjoint; however,
$\vert \vert L_m v \vert \vert$
is infinite for a generic $v \in \mbox{\gotic F}$\\
2.b)For scalar products (~\ref{inners}) in {\goth s},
 such that multiplication by s is just a  bounded operator, not necessarily
a self adjoint one, operators $L_m$ are not self adjoint;
for some scalar products, $\vert \vert L_m v \vert \vert_{\mbox{\gotic F}} $ is
finite for a generic $v \in \mbox{\goth F}$.

\section{ $ L_0$ vs. Loop Equation}

Let us modify  the construction of the representation, (~\ref{rule}) and
(~\ref{scalarF}), as follows: let
$ \mbox{\goth A} = \bigoplus_n \mbox{\gotic H}^n$ (no symmetrization, unlike
the previous construction). The inner
product in {\goth A} is defined as
\bge
\bga
\lt \eta_1 \otimes \eta_2 \ldots  \otimes \eta_n ,
\xi_1 \otimes \xi_2 \ldots \otimes \eta_m  \rt = \\
= \left\{
         \bga  0,\ \ \  m\neq n \\
           \dst\prod_{i=1}^{n}  \lt \eta_{i} , \xi_{i} \rt ,\ m=n
         \ena
  \right.
\ena \label{scalarA}
\ene
In this new representation,  currents act on  $v= \eta_1 \otimes \eta_2 \otimes
\ldots \otimes \eta_n \in \mbox{{\goth A}}$,
with  $\{ \eta_i \}  \in \mbox{\goth H}$, as follows:
\bge
\bga
T[I  (f(s))] . v  = \Bigl( D[I[f(s)]] + \be [I[f(s)]] -\  \lt \be [I[{f(s)}^*
\ 1]], \Bigr). v \equiv \\
\equiv \dst\sum_{i}    \eta_1 \otimes \eta_2 \otimes \ldots \otimes \bigl(
D[I[f(s)]]\eta_i  \bigr)
\ldots \otimes \eta_n + \\
+ \dst\sum_{i=0}^{n}  \eta_1 \otimes \eta_2 \otimes \ldots \eta_i \otimes \be
[I[f(s)]] \otimes \ldots \otimes \eta_n - \\
- \dst\sum_{i} \lt \be [I[f(s)^* 1]], \eta_i \rt\  \eta_1 \otimes \eta_2
\otimes \ldots \eta_i \mbox{\hspace{-4.5 mm} /
 \hspace{1 mm}} \ldots \otimes \eta_n
\ena \label{ncrule}
\ene
It is easy to check that the commutation relations are satisfied, and that
operators $T[I  (f(s))]$
are scew self adjoint for scalar products in {\gotic s} where multiplication by
$s$ is self-adoint.
We construct operators $L_m$  in the universal enveloping algebra as before,
(~\ref{l}).

There are similarities between  two infinite-dimensional Laplacians,
$ L_0 $ and the loop operator {\goth L} for a Wilson loop (monodromy )
$W[x(s)]$ of  the Yang-Mills connection $A(x) dx$ on $S^n$,
$
W[x(s)]= \dst\int DA(x) \ P\exp\left(i \oint_{x(s)} \sum_{1}^{4}
A_\mu dx^\mu\right)\exp\left(-\dst\int_{\bf{S}^n} F\wedge \ast F \right),
$
with the loop equation $ \mbox{\goth L}\  W[x(s)]\equiv
\dst\int ds_1 ds_2 \delta(s_1 -s_2) \dst\frac{\delta}{\delta x(s_1)}
\dst\frac{\delta}{\delta x(s_2)} W[x(s)] =0;
$
for a non-selfintersecting path $\{ x(s) \}$  being a consequence of the  Yang
Mills equation for the connection $A(x) dx$.

Let us think about $v\ \in \mbox{{\goth A}}$ as about  certain  functions in
many variables,\\
$v(X_{(1)}, t_1,X_{(2)}, t_2,
X_{(n)}, t_n )  = \eta_1 (X_{(1)}, t_1) \otimes \eta_2 (X_{(2)}, t_2)  \otimes
\ldots \otimes \eta_n (X_{(n)}, t_n)$,
where each $X_{(a )}$ takes values  in $S^n$. Then
\bge
\bga
L_0 . v = \\ \Bigl( \dst\sum_{k= - \infty}{\infty}
 \dst\sum_{a=1}^{n} \Bigl( \dst\sum_{i<j}   D_{X_{(a )}}[I_{ij}(e[-k](t_a) )]\
 D_{X_{(a)}} [I_{ij}( e[k](t_a) ) ] \\
\hspace{50mm} - \dst\sum_{i}  D_{X_{(a )}}[I_{i,0}( e[-k] (t_a))] \   D_{X_{(a
)}}[I_{i,0}(e[k](t_a))] \Bigr) v +
\\[10mm]
+\dst\sum_{a \neq b \  =1}^{n} \Bigl( \dst\sum_{i<j}   D_{X_{(a
)}}[I_{ij}(e[-k](t_a) )]\
 D_{X_{(b)}} [I_{ij}( e[k](t_b) ) ]- \\
\hspace{50mm} - \dst\sum_{i}  D_{X_{(a )}}[I_{i,0}( e[-k] (t_a))] \
D_{X_{(b)}}[I_{i,0}(e[k](t_b))] \Bigr) v \Bigr)\\[6mm]
+ \mbox{(terms with not more then one derivative)},
\ena \label{la}
\ene
where $D_{X_{(a )}}[I(f(t_a)]$ are  differential operators     acting  as
differentiation only in $X_{(a )}$ variables,
with coefficients which are functions of $X_{(a )}, t_a$

The terms in (~\ref{la}) which involve 2 differentiations in the same variable
cancel out, since in fact the second order
operator restricted  to diagonals ( we will call this restricted operator
$\tilde{\mbox{\gotic L}}_a$, $a=1,2,\ldots n$ )is zero,
\bge
\bga
\tilde{\mbox{\gotic L}}_a := \dst\sum_{k=-]infty}^{\infty} \Bigl(
\dst\sum_{i<j}   D_{X_{(a )}}[I_{ij}(e[-k](t_a) )]\
 D_{X_{(a)}} [I_{ij}( e[k](t_a) ) ] -  \\
\hspace{50mm} - \dst\sum_{i}  D_{X_{(a )}}[I_{i,0}( e[-k] (t_a))] \   D_{X_{(a
)}}[I_{i,0}(e[k](t_a))] \Bigr)
= 0
\ena \
\label{loopl}
\ene
in {\goth H} for every $a$. We view this equation as in some sence a discrete
approximation to the loop equation. The restricted operator $\mbox{\gotic L}_a$
consists of 2 pieces, the spin operator for the maximum compact
subgroup of rotations of $S^n$, and the non-compact piece; they both come with
the same infinite constant $\dst\sum_{k = -\infty}^{\infty} 1 $  (which in some
regularization, is pretty small, $ 1 + 2 \zeta (0) = 0 $ ) ; but  since the
constant
is the same for both terms, we can divide by this constant to compare. Dividing
by this constant, we will get
$$
\bga
\mbox{\gotic L}_a :=   \Bigl( \dst\sum_{i<j}   D_{X_{(a )}}[I_{ij}(1)]\
 D_{X_{(a)}} [I_{ij}( 1 ) ] -  \\
\hspace{50mm} - \dst\sum_{i}  D_{X_{(a )}}[I_{i,0}( 1)] \   D_{X_{(a
)}}[I_{i,0}(1)] \Bigr)
= 0
\ena
$$
The spin operator $\dst\sum_{i<j}   D_{X_{(a )}}[I_{ij}(1)]\
 D_{X_{(a)}} [I_{ij}( 1 ) ] $ is unitary and compact; in some basis it reduces
to multiplication by constants, the
total spin. This spin cannot be equal to zero, since the only eigenvalue for
zero spin is the function
which is identically one, and those were thrown out our Hilbert space {\gotic
h}. The remaining piece is the usual laplacian operator, up to lower order
terms in derivatives,  in a tangent hyperplane to a sphere;
thus what we observe is that restrictin of $L_0$ to the  diagonals yields that
the  states in \mbox{\goth A} are eigenfunctions  of the restricted to the
diagonal laplacian
\bge
\dst\sum_{i}  D_{X_{(a )}}[I_{i,0}( 1)] \   D_{X_{(a )}}[I_{i,0}(1)],
\ene
which is just a usual laplacian modulo terms with lower number of  derivatives.
Thus we obtain the following
\prp
1) the reduction of $L_0$ in {\goth A} to the diagonals is identically zero \\
2) the above is equivalent to the fact that the restriction of the usual
laplacian to the diagonals is zero
on the subspace in {\gotic h}  of solutions of the equation, given by the spin
operator plus lower in derivative terms

\section{Conclusions} We   constructed canonical non-highest weight unitary
irreducible
representation of $\widehat{so}(1,n)$ current algebra as well as canonical
non-highest weight non-unitary representations,
We constructed certain Laplacian operators in the universal enveloping algebra,
acting in representation space. We speculated about possible relation of those
Laplacians with the loop operator for the Yang-Mills.

\newpage

\section{Acknowledgments}
Conversations with A. Vershik, I.M. Gelfand,   B. Bakalov,
are appreciated. I am grateful to IHES for a very stimulating atmosphere, the
financial support,
and hospitality.

\end{document}